\documentclass[12pt]{article}
\pdfoutput=1
\usepackage{color}

\usepackage{jcappub}
\usepackage{amsthm}

\usepackage{multirow} 
\usepackage{bm}
\usepackage{amsmath}
\usepackage{color}
\usepackage{natbib}
\usepackage{graphicx} 

\makeatletter
\renewcommand\footnoterule{%
  \vspace{-5pt}
  \kern-3\p@\hrule\@width.4\columnwidth%
  \kern10\p@}
\makeatother

\def\be{\begin{equation}}
\def\ee{\end{equation}}
\def\ba{\begin{eqnarray}}
\def\ea{\end{eqnarray}}

\newcommand{\Mpl}{M_{\rm Pl}}

\newcommand{\fnl}{f_{\mathrm{ NL}}}

\definecolor{darkgreen}{cmyk}{0.85,0.2,1.00,0.2} 
\definecolor{purple}{cmyk}{0.5,1.0,0,0}

\newcommand{\ModeCode}{{\sc Mode\-Code}}
\newcommand{\MultiNest}{{\sc MultiNest}}
\newcommand{\CAMB}{{\sc CAMB}}
\newcommand{\CosmoMC}{{\sc CosmoMC}}

\begin{document}
\title{Constraining Monodromy Inflation}

\author[1]{Hiranya V. Peiris}
\author[2]{Richard Easther}
\author[3,4]{Raphael Flauger}

\affiliation[1]{Department of Physics and Astronomy, University College London, \\London WC1E 6BT, U.K.}
\affiliation[2]{Department of Physics, University of Auckland, Private Bag 92019, \\ Auckland, New Zealand}
\affiliation[3]{School of Natural Sciences, Institute for Advanced Study,\\ Princeton, NJ 08540, USA}
\affiliation[4]{Center for Cosmology and Particle Physics, Department of Physics,\\
New York University, New York, NY, 10003, USA}
\emailAdd{h.peiris@ucl.ac.uk}
\emailAdd{r.easther@auckland.ac.nz}
\emailAdd{flauger@ias.edu}

\date{\today}

\abstract{ 
We use cosmic microwave background (CMB) data from the 9-year WMAP release to derive constraints on monodromy inflation, which is characterized by a linear inflaton potential with a periodic modulation. We identify two possible periodic modulations that  significantly improve the fit, lowering $\chi^2$ by approximately $10$ and $20$.  However, standard Bayesian model selection criteria assign roughly equal odds to the modulated potential and the unmodulated case.   A modulated inflationary potential can generate substantial primordial non-Gaussianity with a specific and characteristic form. For the best-fit parameters to the WMAP angular power spectrum, the corresponding non-Gaussianity might be detectable in upcoming CMB data, allowing nontrivial consistency checks on the predictions of a modulated inflationary potential. 

}

\maketitle

\section{Introduction} \label{sec:intro}

Monodromy inflation is a theoretically well-motivated scenario for implementing  a period of nearly exponential expansion in the very early universe~\cite{Silverstein:2008sg,McAllister:2008hb,Flauger:2009ab,Berg:2009tg}. Derived within string theory, monodromy models ``unwrap'' angular directions in field space, and have monomial potentials over super-Planckian distances. The underlying periodicity of the field may be imprinted on the potential in the form of a small oscillatory modulation.   The evolving inflaton is often described as a clock that coordinates the inflationary expansion over super-Hubble scales.  The periodicity of the compact direction adds another timescale to the problem -- so that the metaphorical inflaton clock now has both an hour hand and a second hand. 

Phenomenologically, monodromy inflation has a rich set of observable signatures.  In particular, the Lyth bound~\cite{Lyth:1996im} implies that the super-Planckian field ranges are associated with a detectable primordial gravitational wave background. The modulated potential naturally leads to an oscillatory power spectrum, which is a non-trivial signature of a monodromy-driven inflationary era. Moreover, monodromy inflation provides a stringy implementation of resonant non-Gaussianity~\cite{Chen:2008wn,Flauger:2010ja}, generating a 3-point function whose properties would be strongly correlated with the power spectrum. If couplings to other degrees of freedom are small, monodromy potentials have the ability to generate oscillons -- long lived, localized, pseudo-stable excitations of a scalar field -- in the post-inflationary universe~\cite{Amin:2011hj}. Lastly, couplings of the inflaton to a gauge field, when present, may also lead to interesting phenomenology~\cite{Barnaby:2011qe}.

We determine the observational constraints from the cosmic microwave background [CMB] on the canonical model of monodromy inflation, using the 9-year Wilkinson Microwave Background Anisotropy Probe [WMAP] dataset \cite{Bennett:2012fp,Hinshaw:2012fq}. We estimate the likely amplitude of the 3-point signal associated with the parameter values that best fit the angular power spectrum data. The specific form of this signal is not well captured by standard measures of non-Gaussianity, but can be recovered from CMB data with a suitable estimator, such as a modification of the modal expansion method~\cite{Fergusson:2010dm} (see, e.g., Ref.~\cite{Meerburg:2010ca}). We show that information from the 3-point function could  tighten constraints on any possible modulation.   

Our analysis uses the \ModeCode\ package\footnote{ \ModeCode\ can be downloaded from {http://zuserver2.star.ucl.ac.uk/$\sim$hiranya/ModeCode/ } and applies a patch to \CosmoMC. For further details and instructions consult the documentation supplied with the code. }  
\cite{Mortonson:2010er,Easther:2011yq,Norena:2012rs}, an extension to \CosmoMC\ \cite{Lewis:2002ah} developed by two of us [RE and HP] and other collaborators.  \ModeCode\ solves the inflationary perturbation equations numerically  using an algorithm originally developed in Ref.~\cite{Adams:2001vc}, avoiding any analytical approximations beyond those inherent in the underlying cosmological perturbation theory.  This approach allows us to directly estimate the free parameters in the inflationary potential, independently of any empirical characterizations of the primordial power spectrum. Separately, \ModeCode\ replaces the usual Markov Chain sampler in \CosmoMC\ with  the nested sampler \MultiNest\ \cite{Feroz:2007kg, Feroz:2008xx}. This  copes naturally with multimodal likelihood surfaces, which  can arise with modulated potentials or ``feature'' models, and simultaneously computes the Bayesian evidence (e.g., Ref.~\cite{Jaynes2003}).

Calculating the Bayesian evidence addresses the model selection problem,  determining whether the improved fit to the data that follows from inclusion of the modulations  -- or any other  parameters we might add -- justifies the additional complexity of the resulting model. We  compare the modulated potential to the standard monodromy potential without modulations, finding that the usual model selection criteria assign slightly better odds to the modulated model, despite the fact that the former has three extra parameters.

This paper is organized as follows. In Section~\ref{sec:monod} we summarize the properties of monodromy inflation and specify the priors for the parameters in the potential. In Section~\ref{sec:results} we present the parameter estimates and Bayesian evidence values. Section~\ref{sec:3point} discusses the expected 3-point statistics that are permitted by the constraints derived for the potential, and we discuss our results in Section~\ref{sec:discuss}.

\section{Monodromy Inflation, Perturbations and Methodology\label{sec:monod}}

\subsection{Potential and Model}

We work with the monodromy potential 
\be
V(\phi) = \mu^3 \left[\phi - bf  \left( \cos\left(\frac{\phi}{f} + \psi \right) -c \right)  \right]  \label{eq:v}  \, .
\ee 
Here $\mu$ and $f$ are energy scales, while $b$ is a dimensionless parameter and we include a constant phase $\psi$ as in Ref.~\cite{Flauger:2009ab}. 
The linear term in the potential corresponds to the monodromy, and the superimposed oscillatory term is the characteristic potential of axions first considered in the context of inflation in Ref.~\cite{Freese:1990rb}. For convenience, we have introduced an additional dimensionless parameter $c$ and choose it so that the potential vanishes at $\phi=0$. This potential  describes  the inflationary phase but  is unbounded below, and thus does not describe the reheating physics. We assume that $\phi$ is initially large and positive.  

This potential is simplified relative to the ``full'' monodromy potential in Ref.~\cite{McAllister:2008hb}, which has a quadratic minimum at the origin.  The version here does  not significantly modify the potential in the region relevant to the cosmological perturbations, but does change the field value at which inflation ends.   

The equations of motion obeyed by the field, scale factor and perturbations as implemented in~\ModeCode\ are described in Refs.~\cite{Mortonson:2010er,Easther:2011yq,Norena:2012rs}. We will need to refer to the number of $e$-folds of inflation
\be \label{eq:nval}
N= \int_{\phi_{end}}^\phi \frac{d\phi}{\sqrt{2\epsilon}\Mpl} \qquad  \Rightarrow\qquad  N \approx \frac{\phi^2}{2\Mpl^2}  \,.
\ee
The approximate equality is a reliable estimate for moderate values of $b$. For $N=55$, $\phi$ is a little larger than $10 \Mpl$ and between  (say) $N=55$ and $N=45$, $\delta \phi \sim  \Mpl$.   Setting aside the physical constraints on $f$ described below, if $f\gtrsim 10^{-1}\Mpl$ the portion of the potential which generates perturbations on astrophysical scales will resemble an unmodulated potential with a ``running'' spectral index (see also Ref.~\cite{Kobayashi:2010pz}) while if $f\ll \Mpl$ the modulations will cause the power spectrum to vary on a scale that is very short compared to all the other variables in the problem.

\subsection{\ModeCode, Power Spectrum and Evidence}  

The monodromy power spectrum can be computed via an approximate but accurate analytic calculation \cite{Flauger:2009ab,Flauger:2010ja} or numerically via the routines in \ModeCode, which easily accommodates new potentials. We worked with the numerical calculation as it   remains valid  even if the model parameters take extreme values. In order to accurately compute the power spectrum in the small $f$ limit, the initial conditions for the mode evolution are set further inside the horizon than is required for other models, and we significantly boosted the  sampling of the primordial power spectrum $P(k)$ and the accuracy settings within \CAMB, relative to their default values.

 \ModeCode\ incorporates the \MultiNest\  sampler and  computes the {\em evidence\/} $E$, the integrated likelihood $\mathcal{L}$ over the parameter volume $\{\alpha_1,\cdots,\alpha_M\}$, 
\begin{equation} \label{eq:Evidencefull}
E=\int d\alpha^M P(\alpha_i)\mathcal{L}(\alpha_i)\, ,
\end{equation}
weighted by the prior $P(\alpha_i)$,  normalized so that $\int  d\alpha^MP(\alpha_i) \equiv 1$.  For  uniform priors  $P(\alpha_i)$ is a constant (or zero),  and the weighting is the inverse parameter volume, or
\begin{equation}\label{eq:flatprior}
E=\frac{1}{\text{Vol}_M}\int d\alpha^M \mathcal{L}(\alpha_i)\, .
\end{equation}
Evidence allows comparisons between two models, rather than an absolute measure of quality.    Our comparison model is inflation with a purely linear potential  $V(\phi) = \mu^3 \phi$.  The computed evidence  depends on the parameter ranges allowed by the priors, as discussed in detail in Ref.~\cite{Easther:2011yq}.

\begin{table}[tb]
\begin{center}
\begin{tabular}{|l|c|}
\hline
\hline
\multicolumn{2}{|c|}{Inflation}   \\
\hline 
\hline
Mass scale & $-3.615 < \log_{10}(\mu/\Mpl) < -3.015$    \\ \hline
Axion decay constant & $-3.4 <\log_{10}(f/\Mpl)<-2.0$  \\ \hline
Oscillation amplitude & $0<  b<0.9$  \\ \hline
Phase & $-\pi < \psi < \pi$ \\
 \hline
\hline
%
\multicolumn{2}{|c|}{Matching} \\
\hline 
\hline
$e$-foldings &   $N=55$  \\
\hline
\hline
%
%
\multicolumn{2}{|c|}{Astrophysics}   \\
\hline 
\hline
Baryon fraction  &$0.0218859 <  \Omega_\mathrm{b} h^2 < 0.02378859$ \\
 \hline
Dark matter  &  $ \Omega_\mathrm{dm} h^2 =0.1145$  \\
\hline
Reionization & $\tau =0.0874$ \\
\hline
Projected acoustic scale & $\theta =1.040 $ \\
\hline
Sunyaev-Zel'dovich Amplitude &  $A_{\rm SZ} = 0.10078 $   \\
\hline
\hline
\end{tabular}
\end{center}

\caption{Parameter priors used in the analysis. For linear inflation, $b=0$ and the $f$, $\psi$ parameters are unused.   We assume that $\Omega_k$ is zero in all cases, and  parameters are drawn from a uniform prior.   The pivot scale  is $k_\star = 0.05$~Mpc$^{-1}$.  The non-primordial parameters except for  $\Omega_\mathrm{b} h^2$ are fixed at their maximum likelihood values for the unmodulated case.}
\label{tab:priors}
\end{table}

\subsection{Model Specification and Priors}

The energy scale of thermalization, and thus the ``average'' equation of state in the post-inflationary universe, is largely unconstrained -- there is no hard evidence that the universe is thermalized above   MeV scales at which neutrino freeze-out occurs.    The mapping between present day scales and the value of $N$ at which the corresponding mode leaves the horizon and  the value of $\phi$ at horizon crossing is thus poorly constrained \cite{Adshead:2010mc,Easther:2011yq}. Consequently, a (very) small change in the post-inflationary equation of state is degenerate with the phase variable in the potential; therefore, in what follows, we will fix the number of $e$-folds at which the pivot scale leaves the horizon to $55$.

A preliminary investigation of this system varying both the standard cosmological parameters and the free parameters in the potential showed that the posterior on $\log_{10}{(f/\Mpl)}$  has a peak near $f\sim \Mpl$, approaches zero near $\log_{10}{(f/\Mpl)} \sim -1.5$ and rises substantially at smaller values of $f$. Given that values of $f \gtrsim 10^{-1}\Mpl$ are inconsistent with the assumptions under which the monodromy potential was derived, it is clear that we only need to concern ourselves with a ``fast'' modulation of the power spectrum in the small $f$ limit.  Consequently, we can perform our analysis after fixing the usual cosmological parameters, except for $\Omega_\mathrm{b} h^2$ which potentially exhibits a slight degeneracy with $f$ (see below).  The specific values we use to fix the remaining parameters came from the an initial run performed without modulations,  and the priors are presented in Table~\ref{tab:priors}.

Physically, only  weak constraints can be placed on the free parameters in equation~(\ref{eq:v}).  The value of $\mu$   sets the amplitude of the primordial perturbation spectrum, but is not predicted by fundamental theory.  As described in Ref.~\cite{Easther:2011yq}, \ModeCode\ excludes parameters for which the power spectrum amplitude is incompatible with basic structure formation requirements, and this stipulation  dominates the permitted range of $\mu$  when it is specified with a logarithmic prior.  The axion decay constant $f$  is also an unknown mass scale with a large possible range. On very general grounds we expect $f \lesssim  \Mpl$~\cite{Banks:2003sx,ArkaniHamed:2006dz},  and the approximations underlying the derivation of the modulated potential become unreliable if $f \gtrsim 10^{-1}\Mpl$ \cite{McAllister:2008hb,Flauger:2009ab}. For a small range $f \gtrsim 10^{-2}\Mpl$ there is a degeneracy between $f$ and $\Omega_\mathrm{b}$~\cite{Flauger:2009ab}. To avoid such potential degeneracies with the standard cosmological parameters, we work with $f \le 10^{-2}\Mpl$, and we vary $\Omega_\mathrm{b} h^2$ in order to eliminate any effects arising from this degeneracy. We can then self-consistently fix the remaining parameters to their best-fit values. The validity of the approximations made in Refs.~\cite{McAllister:2008hb,Flauger:2009ab} also imposes a lower bound on $f$, and we use $f\gtrsim 4\times10^{-4}\Mpl$. (A similar bound can also be derived by requiring weak coupling of the effective theory~\cite{Behbahani:2011it}.) In order to avoid potential numerical singularities, we work with the limit $b<0.9$.
 
\section{Parameter Estimates and Evidence \label{sec:results}}

\begin{figure}[tb]
\begin{center}
\includegraphics[scale=0.52]{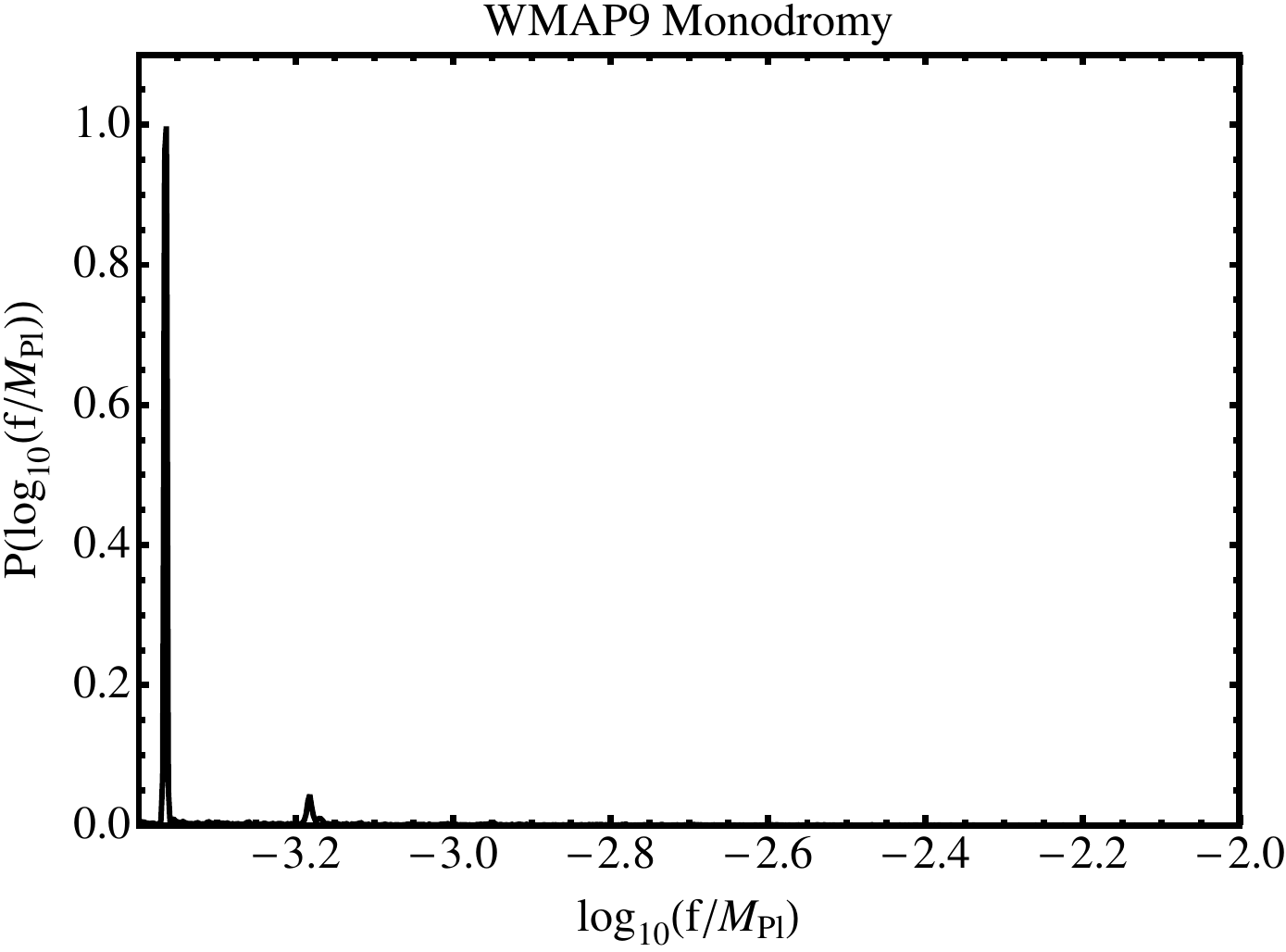}
\includegraphics[scale=0.5]{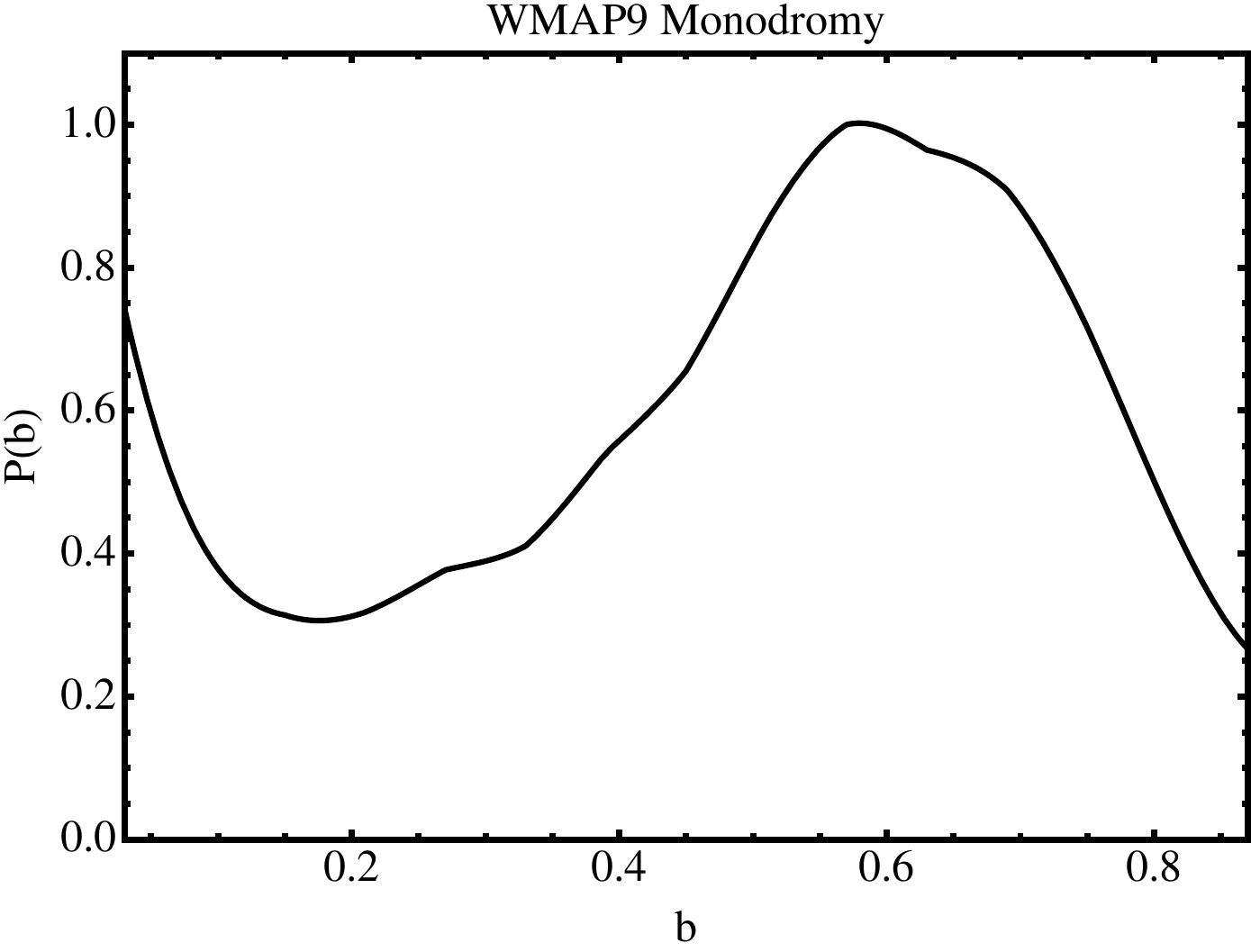}  
\end{center}
\caption{\label{fig:1D} Marginalized posterior distributions for inflationary parameters.}
\end{figure}

We base our constraints on the WMAP 9-year dataset \cite{Bennett:2012fp,Hinshaw:2012fq}. Constraints on the monodromy model from WMAP7 were derived in Refs.~\cite{Meerburg:2011gd,Aich:2011qv}, and other studies of oscillatory features in the CMB include Refs.~\cite{Martin:2003sg,Martin:2004iv,Martin:2004yi,Chen:2012ja}. The Atacama Cosmology Telescope [ACT] \cite{Sievers:2013wk}   and South Pole Telescope [SPT] \cite{Hou:2012xq}  datasets contain information on the power spectrum at angular scales unresolved by WMAP. However,  the ACT and SPT likelihoods are binned into bandpowers,  limiting their sensitivity to short wavelength modulations. Moreover, the two datasets lead to apparently different conclusions about the form of the power spectrum in the smooth limit \cite{DiValentino:2013mt}. Consequently, we do not include either of them in our analysis.
Modulations in the primordial power spectrum also lead to modulations in the galaxy power spectrum measured from large scale structure surveys. Currently these bounds are weaker\footnote{See however Ref.~\cite{Huang:2012mr} for a Euclid forecast.} and we focus on the WMAP data.

\begin{figure}[tb]
\begin{center}
\includegraphics[width=6in]{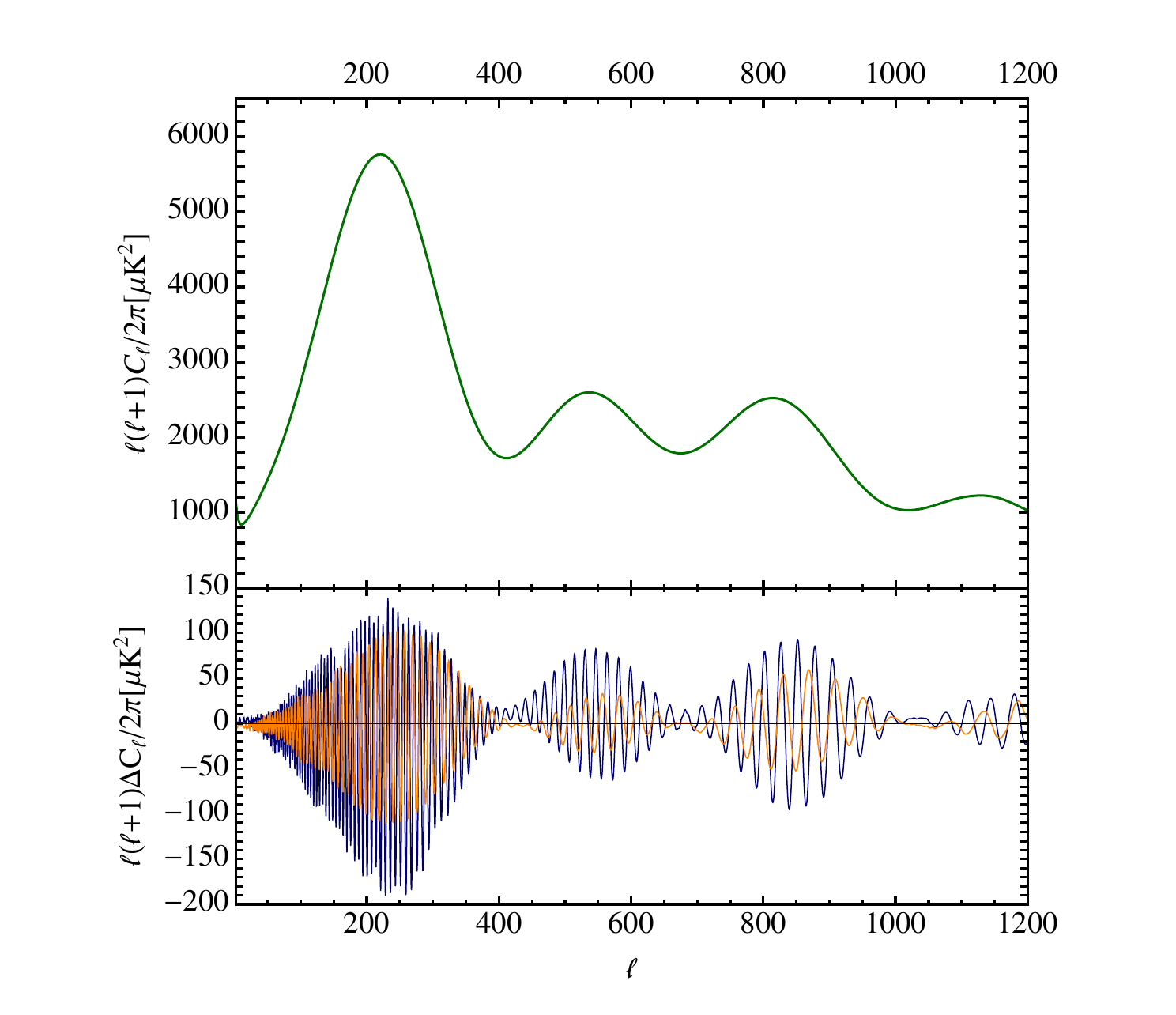}
 \end{center}
\caption{\label{fig:cl}   Reference angular power spectrum with $b=0$ and difference for the best fit spectrum with $\log_{10}{(f/\Mpl)}\approx -3.38$ in blue and with $\log_{10}{(f/\Mpl)}\approx -3.18$ in orange.} 
\end{figure}   
  
Figure~\ref{fig:1D} shows  posterior distributions for the free parameters $f$ and $b$.  
Two spikes in the posterior for $\log_{10}{(f/\Mpl)}$ are  visible, and the improvement in the effective $\chi^2$ or $-2 \log \mathcal{L}$ associated with the lefthand peak ($\log_{10}{(f/\Mpl)}\approx -3.38$) is approximately $\Delta \chi_\mathrm{eff}^2 = 19.7$ over the unmodulated potential, and $\Delta \chi_\mathrm{eff}^2 = 18.9$ with respect to $\Lambda$CDM, which has best-fit $\chi_\mathrm{eff}^2 =  7558.0$ for WMAP9.  The posterior for $b$ contains a peak away from $b=0$ associated with the two likelihood peaks; however, the data do not exclude $b=0$, and the marginalized posterior probability exhibits another rise there. We do not plot the posterior for the phase $\psi$, as it is almost uniform, essentially reproducing the prior. This is consistent with the peaks seen in $\log_{10}{(f/\Mpl)}$, as the marginalized posterior for the phase is not dominated by the narrow peaks in $\log_{10}{(f/\Mpl)}$. 

The righthand peak at $\log_{10}{(f/\Mpl)} \approx -3.18$ leads to an improvement $\Delta \chi_\mathrm{eff}^2 = 11.7$ with respect to the unmodulated potential. It is consistent with the best-fit point  in Ref.~\cite{Flauger:2009ab}, where it had an improvement in the $\Delta \chi_\mathrm{eff}^2$ of $11$.   Ref.~\cite{Flauger:2009ab} also finds a spike in the WMAP5 likelihood which is consistent with the higher peak seen here,  in  both frequency and amplitude.   Ref.~\cite{Flauger:2009ab} also sees several smaller peaks, and these are less prominent in the new analysis. The two most prominent examples thus ``survive'', albeit after trading roles. The analysis in Ref.~\cite{Flauger:2009ab} made use of the analytic expression for the primordial power spectrum, which provides a non-trivial consistency check for \ModeCode.   When $b$ is increased at fixed $\mu$  and  $f\ll \Mpl$, the averaged value of $P(k)$ increases. Consequently, at large values of $b$, the best fit  spectrum has a slightly smaller value of $\mu$ than the best fit in the $b=0$ limit. Even though we marginalize over  $\Omega_\mathrm{b} h^2$ for robustness, there is no significant degeneracy with this parameter for the monodromy parameter priors studied here.

\begin{figure}[tb]
\begin{center}
\includegraphics[width=6in]{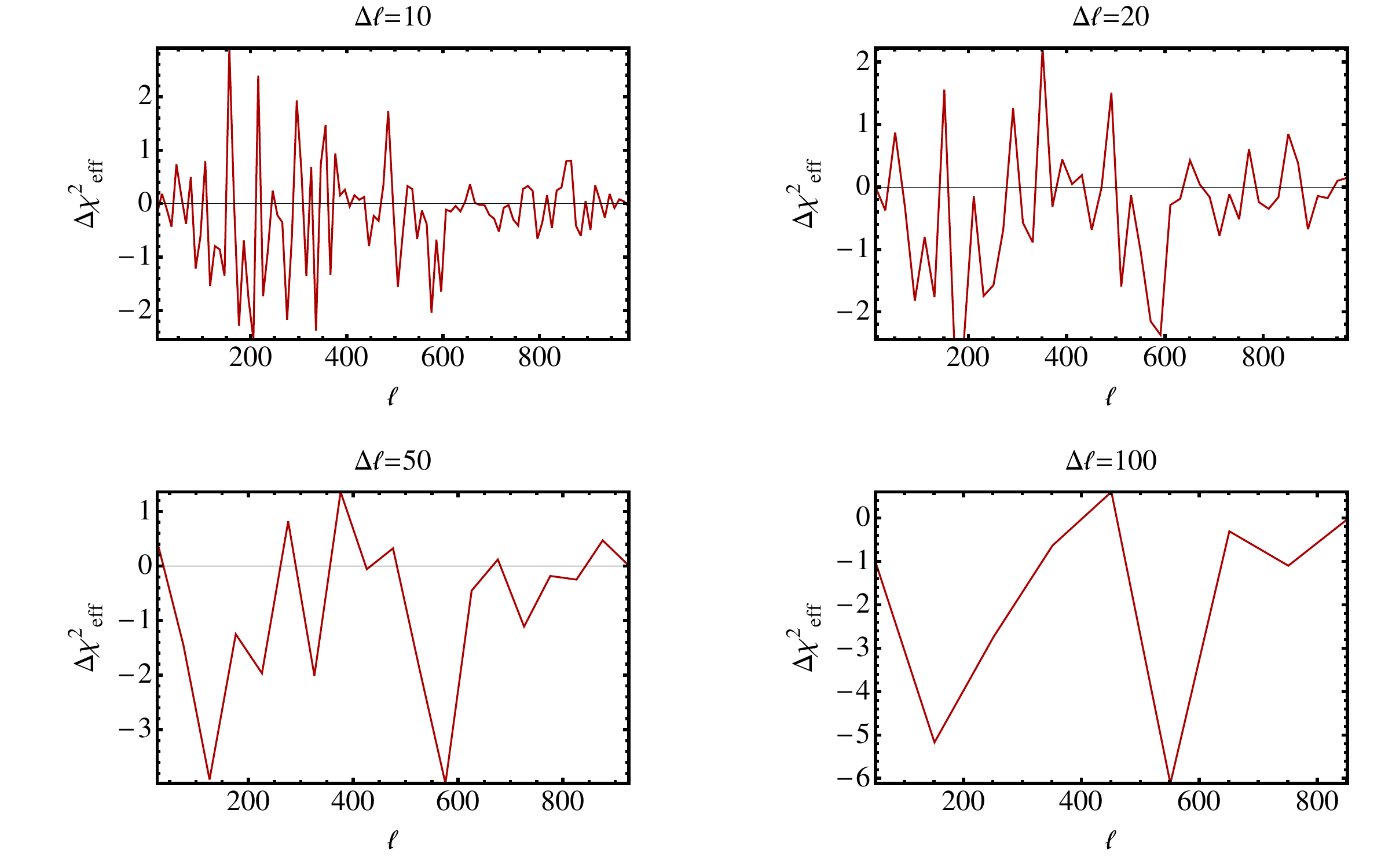}
 \end{center}
\caption{\label{fig:chi2} The difference in $\Delta \chi^2_\mathrm{eff}$ given by the WMAP9 MASTER TTTT likelihood for the reference and best-fit modulated models as a function of $\ell$, binned as indicated by the labels. 
} \end{figure}

A number of other ``feature''  models are known to yield an improvement in $\Delta \chi^2$ of order 10 (e.g. a step in the inflaton potential, yielding high-$\ell$ oscillations in the power spectrum \cite{Adshead:2011jq}) and even if one of them were to be genuine, they cannot all be genuine. However, the enhancement in the fit at the $\log_{10}{(f/\Mpl)}\approx -3.38$ peak is relatively substantial.  We have performed preliminary simulations to determine the significance of such an improvement in a fit to data without a signal, which indicate that the improvement is less significant than it may seem at first sight, in accordance with the results from the model selection analysis.  Further, the approximations in the WMAP likelihood code have not necessarily been tested in the regime of very high frequency oscillations in which our best fit parameters lie.   The $\Delta C_\ell$ for the best-fit parameters associated with the higher peak in the posterior distribution of $\log_{10}(f/\Mpl)$ are shown in Figure~\ref{fig:cl}.    In Figure~\ref{fig:chi2} we show the contributions to $\Delta \chi^2_\mathrm{eff}$ as a function of multipole, for the higher likelihood peak. We see that the reduction in $\chi^2$ comes from the full multipole range, rather than being concentrated in a narrow region, which is consistent with a modulated model.

\begin{figure}[tb]
\begin{center}
\includegraphics[width=6in]{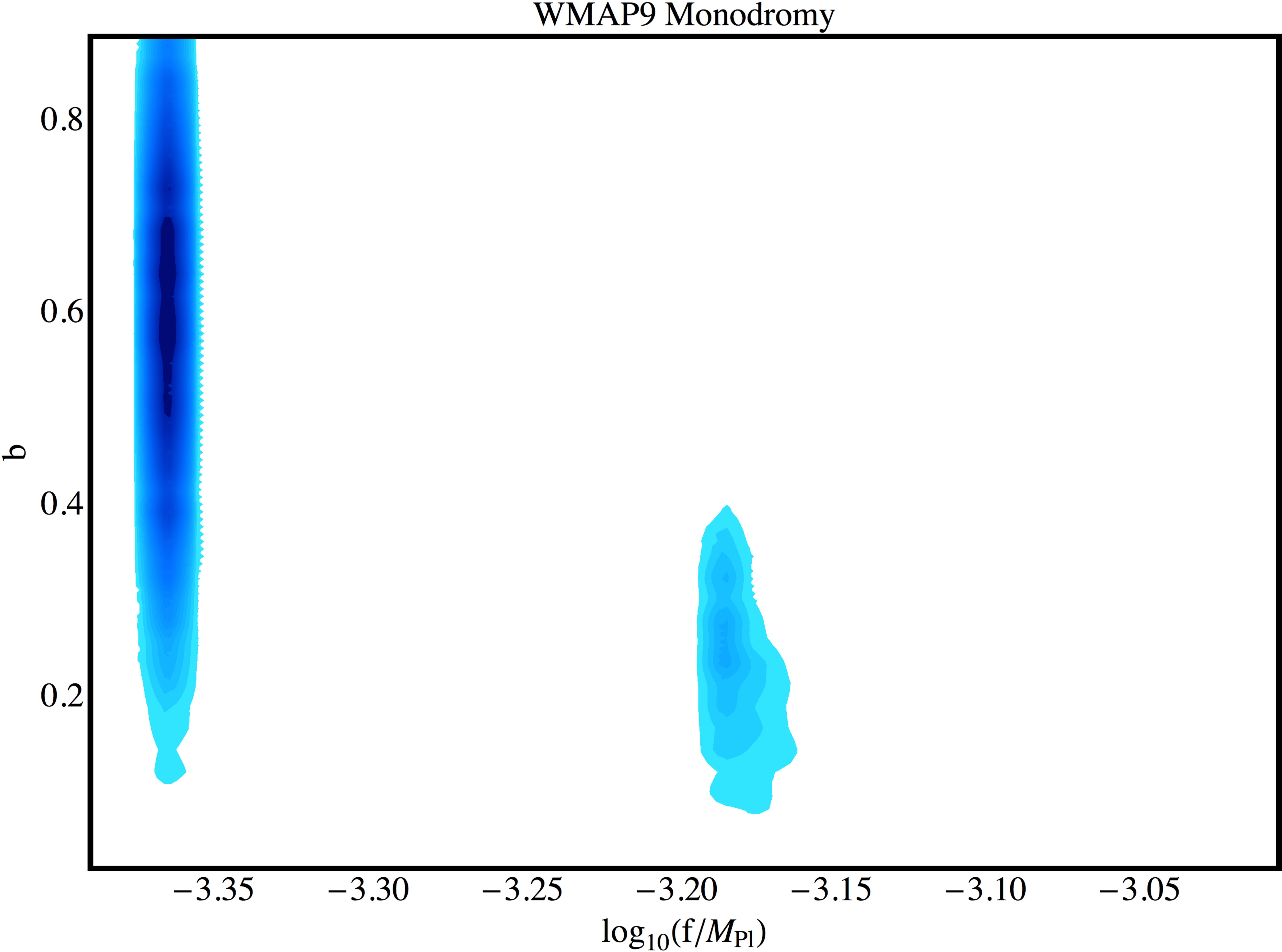}  
 \end{center}
\caption{\label{fig:2D} Marginalized 2D-joint posterior for $b$ and $\log_{10}{(f/\Mpl)}$ -- the intensity is a function of the posterior, and shows that the peaks are narrow and well-defined. Note that the color scale is qualitative. }
\end{figure}

The presence of sharp features in the marginalized posteriors taxes the usual software tools used in cosmological MCMC analyses.  Fortunately, we can make use of the \MultiNest\ sampler \cite{Feroz:2007kg}, which copes with multimodal and rapidly varying likelihood functions.   Plotting  binned 2D posteriors leads to messy results, and we instead show the ``heat map''  of the posterior in Figure~\ref{fig:2D}, mapped to a qualitative intensity scale.    

We compute the evidence ratio between the pure, unmodulated monodromy potential and the modulated case using \MultiNest, finding  $\Delta \log{E} \sim +0.6$ in favor of the latter.
Thus, the betting odds favor the modulated monodromy model by about 2:1 despite its three extra parameters. However, these odds are far from decisive\footnote{Betting odds of 150:1 are considered decisive in this context.}, and the WMAP9 data cannot differentiate between these competing models.

\section{The  Three Point Function \label{sec:3point}}

\begin{figure}[tb]
\begin{center}
\includegraphics[width=4.5in]{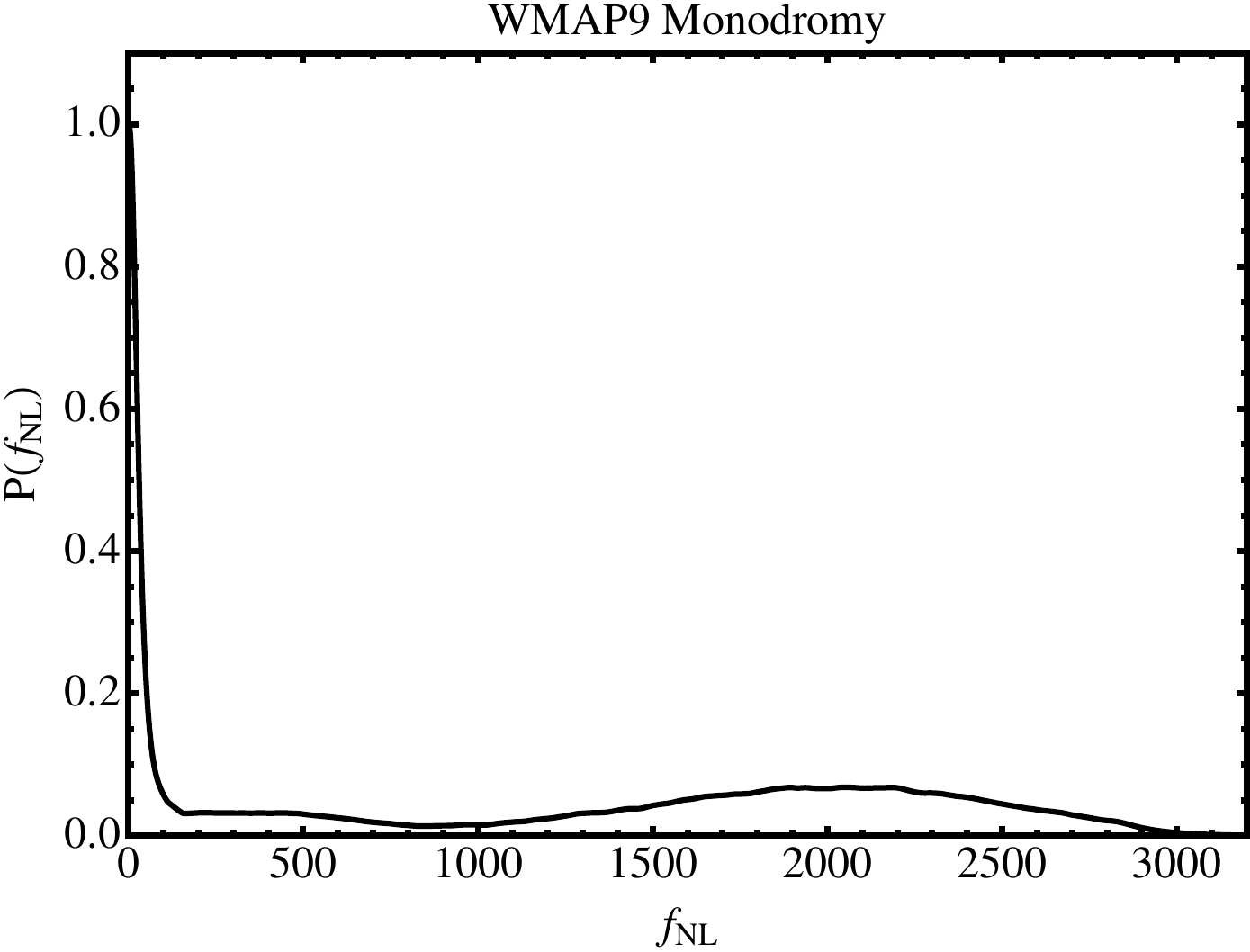}  
 \end{center}
\caption{\label{fig:fnl} Posterior distribution for $f_\mathrm{NL}$. }
\end{figure}

Up until this point much of our analysis is functionally equivalent to working with a modulated primordial power spectrum $P(k)$ (e.g. Ref.~\cite{Easther:2001fz,Easther:2002xe,Martin:2003sg}). However, monodromy inflation is a stringy realization of the resonant non-Gaussianity scenario, first described in Ref.~\cite{Chen:2008wn}, and  makes unambiguous predictions for both the 2-point function and the 3-point function.  Consequently, given our posterior we can also compute the expected amplitude of the   3-point function. The ``shape'' of this non-Gaussianity does not   map directly onto any of the  forms with known constraints, so we cannot use the CMB 3-point function in the parameter estimation process.  However, we can ask whether this information would tighten constraints derived from the 2-point function or, more optimistically,  confirm that a peak seen in the posterior of $\log_{10}{(f/\Mpl)}$ is consistent with an initial period of monodromy inflation.

\begin{figure}[tb]
\begin{center}
\includegraphics[width=5in]{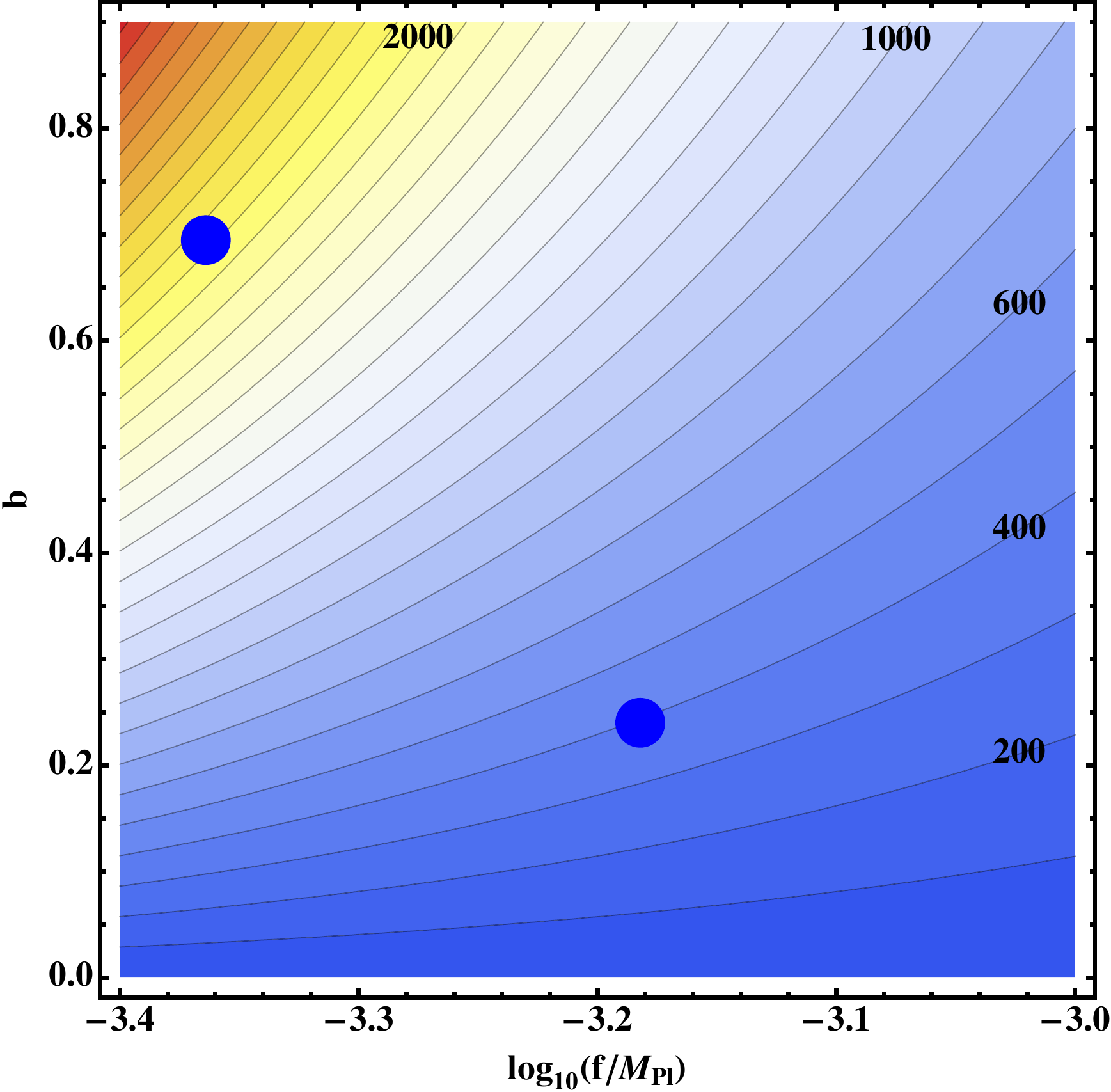}  
 \end{center}
\caption{\label{fig:fnlcontour}  Contours show the predicted amplitude of the resonant non-Gaussianity signal associated with the the modulation parameters $f$ and $b$. The two points show the best fit values associated with the peaks seen in the marginalized distribution for $\log_{10}{(f/\Mpl)}$.  }
\end{figure} 

For this model the approximate amplitude of the 3-point function is \cite{Flauger:2010ja} 
\be \label{eq:fres}
\fnl = \frac{3 \sqrt{2 \pi} b}{8 (f \phi_\star/\Mpl^2)^{3/2}} \, ,
\ee
where $\phi_\star$ is the field value at which the perturbations are generated, which is roughly $\sqrt{2 N} = \sqrt{110}$, given our priors.   We plot the inferred posterior for $\fnl$ in Figure~\ref{fig:fnl} -- this is peaked strongly near zero, but integrating the tail we find that $P(\fnl > 100) \approx 0.75$. 
 
To our knowledge, there are no published constraints on $\fnl$  for resonant non-Gaussianity, or even estimates of the likely constraining power of  datasets such as {\it Planck}\footnote{http://www.esa.int/Planck} for this range of frequencies. However, a simple estimate presented in Ref.~\cite{Behbahani:2011it} indicates that the signal-to-noise ratio for a measurement of the 3-point function is smaller but comparable to the one in the 2-point function in the regime of interest. We plot the predicted value of $\fnl$ as a function of $b$ and $\log_{10}{(f/\Mpl)}$ in Figure~\ref{fig:fnlcontour}, and overlay the best fit points for the two peaks seen in the posterior for $\log_{10}{(f/\Mpl)}$. Resonant non-Gaussianity will be more challenging to constrain than some of the simpler forms.  However,  future data may constrain the power spectrum and $\fnl$  with sufficient accuracy to test whether the peaks seen in the posterior for $\log_{10}{(f/\Mpl)}$ are associated with a modulated inflaton potential. Further motivation for a measurement of the 3-point function comes from models presented in Ref.~\cite{Behbahani:2012be} that can lead to resonant non-Gaussianity without a corresponding signal in the power spectrum. 
  
\section{Discussion  \label{sec:discuss}}

We have used \ModeCode\  \cite{Mortonson:2010er,Easther:2011yq,Norena:2012rs} to derive constraints on monodromy inflation based on the WMAP 9-year dataset.    We find two prominent peaks  in the marginalized distribution for $\log_{10}{(f/\Mpl)}$, the axion decay constant.  Our results are summarized in Table~\ref{tab:summary}.

\begin{table}[tb]
\begin{center}
\begin{tabular}{|l|r|r|}
\hline
\hline
\multicolumn{3}{|c|}{Best-fit parameters}   \\
\hline 
\hline
& peak 1 & peak 2 \\ \hline
Mass scale $\log_{10}(\mu/\Mpl)$ & $-3.22$ & $-3.22$   \\ \hline
Axion decay constant $\log_{10}(f/\Mpl)$  & $-3.36$  & $-3.18$  \\ \hline
Oscillation amplitude $b$ & $0.69$ & $0.24$  \\ \hline
Phase $\psi$ & $2.47$ & $-2.00$ \\ \hline
$\Delta \chi^2_\mathrm{eff}$ relative to unmodulated case & $-19.7$ & $-11.7$ \\
\hline
\hline
\multicolumn{3}{|c|}{Model Comparison}   \\
\hline
\hline
$\Delta \log E$ relative to unmodulated case & \multicolumn{2}{|c|}{$+0.6$}  \\ 
\hline
\end{tabular}
\end{center}

\caption{Summary of results for parameter estimation and model comparison. Best-fit parameters are given for the two peaks seen in the posterior distribution.}
\label{tab:summary}
\end{table}

The improvement in the $\chi^2$ at the peaks is $\sim 12$ and $\sim 20$ with respect to the unmodulated potential, and slightly worse with respect to $\Lambda$CDM.  The concordance cosmological model is a very good fit to the WMAP9 data,  with $\chi^2$ per degree of freedom $1200 / 1168$ for $\ell=33$ -- $1200$, i.e., $\sim 1.03$ \cite{Bennett:2012fp}. Thus, a model which better describes the data has access to an improvement of $\Delta \chi^2_\mathrm{eff} \sim 30$ on average, before it begins over-fitting the data. The higher likelihood peak in the monodromy case gives $\chi^2$ per degree of freedom $1183 / 1168$ in the same multipole range, i.e., $\sim 1.01$. 

The  monodromy model  contains a single modulation, so within the context of our prior, at least one of these peaks is a spurious ``fit to noise'', rather than evidence for a genuine modulation of the primordial power spectrum.  The improvement in fit at the higher peak in the posterior for $\log_{10}{f}$ is noticeable, but not  compelling. Moreover, even if the modulation was present in the processed data, additional work would be required to demonstrate that this was not an artifact of the data analysis.\footnote{The first {\em Planck\/} cosmology data release took place after this work was first posted to the ArXiv. An initial search was made for a modulated power spectrum \cite{Ade:2013uln}, but that scenario does not match the monodromy model considered here.  }  

In future CMB datasets, the polarization is likely to be tightly constrained. The scalar power spectrum dominates the $\langle EE\rangle$ and $\langle TE\rangle$ correlations and so any features in the underlying spectrum will produce a similarly modulated signal in the $E$-mode polarization, allowing much tighter constraints on any putative signal. Moreover, the polarization transfer function is narrower than the temperature transfer function, so features in the primordial spectrum are more clearly resolved in $\langle EE\rangle$  than in $\langle TT\rangle$ \cite{Mortonson:2009qv}.  Beyond improved measurements of the power spectrum, we have also considered the amplitude of the 3-point function associated with  monodromy potential,  demonstrating that this is substantial for parameter values where the likelihood is peaked.  If either of the candidate modulations seen here was confirmed within a higher-quality dataset, the predictions for the 3-point function are likely to be testable. This suggests that after detecting a modulated power spectrum -- which would be an important result in its own right -- it may be possible to determine whether this modulation was associated with a monodromy-style inflaton potential.  Finally, monodromy inflation predicts the existence of a detectable background of gravitational waves, providing a further test of this scenario.

\section*{Acknowledgments}
HVP is supported by STFC, the Leverhulme Trust, and the European Research Council under the European Community's Seventh Framework Programme (FP7/2007-2013) / ERC grant agreement no 306478-CosmicDawn.  RF is supported by the NSF
under grant NSF-PHY-0855425 and NSF-PHY-0645435. We acknowledge the use of the Legacy Archive for Microwave Background Data (LAMBDA). Support for LAMBDA is provided by the NASA Office of Space Science. The authors  acknowledge the contribution of the NeSI high-performance computing facilities and the staff at the Centre for eResearch at the University of Auckland.   New Zealand's national facilities are provided by the New Zealand eScience Infrastructure (NeSI) and funded jointly by NeSI's collaborator institutions and through the Ministry of Business, Innovation and Employment's Infrastructure programme {\url{http://www.nesi.org.nz.}}    The authors are particularly grateful to Gene Sudenkov for his assistance with the  NeSI cluster. This research was supported in part by the National Science Foundation under Grant No. NSF PHY11-25915.

\mbox{}

\bibpreamble{\vspace{-1cm}}



\end{document}